\documentclass[reqno]{amsart}

\usepackage{hyperref}

\newcommand*{\mailto}[1]{\href{mailto:#1}{\nolinkurl{#1}}}
\newcommand{\arxiv}[1]{\href{http://arxiv.org/abs/#1}{arXiv: #1}}
\newcommand{\doi}[1]{\href{http://dx.doi.org/#1}{doi: #1}}

\newtheorem{theorem}{Theorem}[section]
\newtheorem{lemma}[theorem]{Lemma}

\newtheorem{definition}[theorem]{Definition}

\newcommand{\R}{{\mathbb R}}

\newcommand{\C}{{\mathbb C}}

\newcommand{\id}{{\mathrm{id}}}
\newcommand{\loc}{{\mathrm{loc}}}
\newcommand{\cc}{{\mathrm{c}}}

\newcommand{\ran}[1]{\mathrm{ran}(#1)}

\newcommand{\ledot}{\,\cdot\,}

\newcommand{\qd}{{[1]}}
\newcommand{\dip}{\upsilon}
\newcommand{\F}{\mathcal{F}}
\newcommand{\norm}[1]{\left\Vert#1\right\Vert}

\renewcommand{\labelenumi}{(\roman{enumi})}
\renewcommand{\theenumi}{\roman{enumi}}
\numberwithin{equation}{section}

\begin{document}

\title[A Lagrangian view on complete integrability]{A Lagrangian view on complete integrability of the two-component Camassa--Holm system}

\author[J.\ Eckhardt]{Jonathan Eckhardt}
\address{Faculty of Mathematics\\ University of Vienna\\ Oskar-Morgenstern-Platz 1\\ 1090 Wien\\ Austria}
\email{\mailto{jonathan.eckhardt@univie.ac.at}}
\urladdr{\url{http://homepage.univie.ac.at/jonathan.eckhardt/}}

\author[K.\ Grunert]{Katrin Grunert}
\address{Department of Mathematical Sciences\\ Norwegian University of Science and Technology\\NO-7491 Trondheim\\ Norway}
\email{\mailto{katring@math.ntnu.no}}
\urladdr{\url{http://www.math.ntnu.no/~katring/}}

%\thanks{\href{http://dx.doi.org/DOI}{... (to appear)}}
\thanks{{\it Research supported by the Austrian Science Fund (FWF) under Grant No.\ J3455 and by the Research Council of Norway under the grant Waves and Nonlinear Phenomena (WaNP)}}
%\thanks{Research supported by Grant No. {\it J3455} from the Austrian Science Fund (FWF) and by the grant {\it Waves and Nonlinear Phenomena (WaNP)} from the Research Council of Norway}


\begin{abstract}
 We show how the change from Eulerian to Lagrangian coordinates for the two-component Camassa--Holm system can be understood in terms of certain reparametrizations of the underlying isospectral problem. 
 The respective coordinates correspond to different normalizations of an associated first order system.
 In particular, we will see that the two-component Camassa--Holm system in Lagrangian variables is completely integrable as well. 
\end{abstract}

\maketitle

\section{Introduction}

The Camassa--Holm (CH) equation \cite{caho93, cahohy94}
\begin{equation}
u_t-u_{xxt}+3uu_x-2u_xu_{xx}-uu_{xxx}=0,
\end{equation}
which serves as a model for shallow water waves \cite{cola09}, has been studied intensively over the last twenty years, due to its rich mathematical structure. For example, it is bi-Hamiltonian \cite{fofu81}, formally completely integrable \cite{co01}, has infinitely many conserved quantities \cite{le05}, and for a huge class of smooth initial data, the corresponding classical solution only exists locally in time due to wave breaking \cite{coes98, coes98b, coes00}. Especially the last property attracted a lot of attention and led to the construction of different types of global weak solutions via a generalized method of characteristics \cite{brco07, brco07a, hora07, hora09, GHR}. 
For conservative solutions, another possible approach is based on the solution of an inverse problem for an indefinite Sturm--Liouville problem \cite{besasz00, ConservCH, ConservMP, IsospecCH}; the inverse spectral method. 
The aim of this note is to point out some connections between these two ways of describing weak conservative solutions. 

Over the last few years various generalizations of the CH equation have been introduced. A lot of them have been constructed in such a way that one or several properties of the CH equation are preserved. Among them is 
the two-component Camassa--Holm (2CH) system \cite{coiv08}
\begin{subequations}
\begin{align}\label{eqn2CHEul}
 u_t - u_{xxt} + 3uu_x - 2u_x u_{xx} - u u_{xxx} + \rho \rho_x = 0, \\ 
  \rho_t + (u\rho)_x = 0, 
\end{align} 
\end{subequations}
that may also be written in the alternative form 
\begin{subequations}\label{eqn2CHwEul}
\begin{align}
u_t+uu_x+p_x& =0,\\
\rho_t+(u\rho)_x& =0,
\end{align}
where the auxiliary function $p$ solves the differential equation
\begin{equation}\label{eqn2CHPEul}
p-p_{xx}=u^2+\frac12 u_x^2+\frac12 \rho^2.
\end{equation}
\end{subequations}
From our point of view this generalization is of special interest, not only because it has been derived in the context of shallow water waves \cite{coiv08}, but also because weak solutions can be described via a generalized method of characteristics \cite{grhora12, GHR} as well as via an underlying isospectral problem \cite{chlizh06, coiv08, hoiv11}. Thus we are going to study the 2CH system, which reduces to the CH equation when $\rho$ vanishes identically.

As already hinted above and presented in \cite{GHR}, there is not only one class of weak solutions but several of them, dependent on how the energy is manipulated when wave breaking occurs. This means that the spatial derivative $u_x(\ledot, t)$ of the solution $(u(\ledot,t), \rho(\ledot,t))$ becomes unbounded within finite time, while both $\norm{u(\ledot,t)}_{H^1(\R)}$ and $\norm{\rho(\ledot,t)}_{L^2(\R)}$ remain bounded, see e.g.\ \cite{guyi10,guli10, guli11, guzh10,yu12, zhli10} and the references therein. In addition, energy concentrates on sets of measure zero. If one continues solutions after wave breaking in such a way that the total amount of energy, which is described by a Radon measure $\mu$, remains unchanged in time, one obtains the so-called weak conservative solutions \cite{grhora12}. Thus any weak conservative solution is described by a triplet $(u, \rho,\mu)$, where the connection between $u$, $\rho$ and $\mu$ is given through $\mu_{\mathrm{ac}}=(u_x^2+\rho^2)dx$, and $\mu(\R,t)$ is independent of time.

The construction of these solutions by a generalized method of characteristics relies on a transformation from Eulerian to Lagrangian coordinates \cite{grhora12}, based on \cite{brco07, hora07}, which will be reviewed in Section~\ref{secEtoL} (we refer to \cite{grhora12} for details). 
Under this transformation, the 2CH system can be rewritten in Lagrangian variables (for conservative solutions) as 
\begin{subequations}\label{eqn2CHLag}
\begin{align}
 \label{eqn2CHLag1} y_t & = U,\\ 
 \label{eqn2CHLag2}U_t & = -Q, \\ 
 \label{eqn2CHLag3}h_t & = 2(U^2-P)U_\xi, \\ 
 \label{eqn2CHLag4}r_t & = 0,
\end{align}
where the functions $P$ and $Q$ are given by 
\begin{align}\label{intrepP3}
P(\xi,t) & =\frac14 \int_\R e^{-\vert y(\xi,t)-y(s,t)\vert} (2U^2y_\xi+h)(s,t)ds, \\
\label{eqnDefQ} Q(\xi,t) & = -\frac 14 \int_\R \mathrm{sign}(\xi-s)e^{-\vert y(\xi,t)-y(s,t)\vert } (2U^2y_\xi+h)(s,t)ds.
\end{align}
\end{subequations}
Note that the three Eulerian coordinates $(u,\rho,\mu)$ are mapped to four Lagrangian coordinates $(y,U,h,r)$, which indicates that to each element $(u,\rho,\mu)$ there corresponds an equivalence class in Lagrangian coordinates. These equivalence classes can be identified with the help of relabeling functions. 

The purpose of this note is to study what the change from Eulerian to Lagrangian variables means in terms of the isospectral problem underlying the 2CH system. 
It is known  \cite{chlizh06, coiv08, hoiv11} that the 2CH system can be written as the condition of compatibility for the overdetermined system 
\begin{subequations}
\begin{align}\label{eqnIso2CH}
 - \psi_{xx} + \frac{1}{4}\psi  & = z (u-u_{xx}) \psi + z^2 \rho^2 \psi, \\
 \psi_t  & = \frac{1}{2} u_x \psi - \left(\frac{1}{2z}+u\right) \psi_x.
\end{align}
\end{subequations}
In particular, the spectrum associated with~\eqref{eqnIso2CH} is invariant under the 2CH flow. 
We will see in Lemma~\ref{lemEquCanSys} that the isospectral problem, that is, the differential equation~\eqref{eqnIso2CH} can be rewritten as a particular equivalent first order system. 
It then turns out that a normalizing standard transformation (which is well-known in the theory of canonical systems \cite[Section~4]{wiwo12}) takes this system to another equivalent first order system that only involves the Lagrangian variables $(y, U,h,r)$ as coefficients; see Lemma~\ref{lemISOLag}. 
Moreover, relabeling of Lagrangian variables simply amounts to an elementary reparametrization  of the first order system; see Lemma~\ref{lemISOrl}. 
From this point of view, the relation between Eulerian and Lagrangian variables can be understood as different kinds of normalizations of the same (that is, equivalent up to reparametrizations) first order system.  
 Lagrangian coordinates (in $\mathcal{F}_0$) correspond to trace normalization of an associated weight matrix and Eulerian coordinates correspond to normalization of its bottom-right entry. 

That our newly obtained first order system~\eqref{eqnFOSlagrange} indeed serves as an isospectral problem for the Lagrangian version of the 2CH system is then shown in Section~\ref{secCI}. 
More precisely, we will see that the system~\eqref{eqn2CHLag} turns out to be completely integrable in the sense that it can be reformulated as the compatibility condition for an overdetermined system.

\subsection*{Notation} 
  For integrals of a continuous function $f$ with respect to a Radon measure $\nu$ on  $\R$, we will employ the convenient notation 
\begin{align}\label{eqnDefintmu}
 \int_x^y f d\nu = \begin{cases}
                                     \int_{[x,y)} f d\nu, & y>x, \\
                                     0,                                     & y=x, \\
                                     -\int_{[y,x)} f d\nu, & y< x, 
                                    \end{cases} \qquad x,\,y\in \R, 
\end{align}
 rendering the integral left-continuous as a function of $y$. 
 If $f$ is even locally absolutely continuous on $\R$ and $g$ denotes a left-continuous distribution function of $\nu$, then we have the integration by parts formula 
\begin{align}\label{eqnPI}
  \int_{x}^y  f d\nu = \left. g f\right|_x^y - \int_{x}^y g(s) f'(s) ds, \quad x,\,y\in \R,
\end{align}
which can be found in \cite[Exercise~5.8.112]{bo07}  for example.

%%%%%%%%%%%%%%%%%%%%%%%%%%%%%
\section{From Eulerian to Lagrangian coordinates}\label{secEtoL}
%%%%%%%%%%%%%%%%%%%%%%%%%%%%%

In this section we will briefly outline the change from Eulerian to Lagrangian coordinates for the two-component Camassa--Holm system. 
This has been done for the conservative case in \cite{grhora12}, where the interested reader may find additional details.
For the sake of simplicity and readability, we will only consider the case of vanishing spatial asymptotics, that is, when the initial data $(u_0,\rho_0)$ belongs to $H^1(\R)\times L^2(\R)$.
It is well-known that even in the case of smooth initial data, wave breaking can occur within finite time, that is, energy may concentrate on sets of Lebesgue measure zero. 
Dependent on how the concentrated energy is manipulated, one may obtain different kinds of global weak solutions, the most prominent ones being the conservative and dissipative ones; see \cite{GHR}. 
For our purposes (that is, viewing the two-component Camassa--Holm system as an integrable system), the conservative solutions are the appropriate choice.  
 In order to obtain a well-posed notion of global solutions, we need to take wave breaking into account by augmenting the  Eulerian coordinates with a non-negative Radon measure $\mu$ describing the energy of a solution.  

\begin{definition}[Eulerian coordinates]
The set $\mathcal{D}$ is composed of all triples $(u,\rho,\mu)$ such that $u$ is a real-valued function in $H^1(\R)$, $\rho$ is a non-negative function in $L^2(\R)$ and $\mu$ is a non-negative and finite Radon measure on $\R$, whose absolutely continuous part $\mu_{\mathrm{ac}}$ is given by 
\begin{equation}
\mu_{\mathrm{ac}}= \left(u_x^2 +\rho^2\right) dx.
\end{equation}
\end{definition}

The main benefit of the change from Eulerian to Lagrangian coordinates lies in the fact that the measure $\mu$ turns into a function which allows to apply a generalized method of characteristics in a suitable Banach space to solve the two-component Camassa--Holm system in Lagrangian variables. 
Before introducing the set of Lagrangian coordinates $\mathcal{F}$, we have to define the set of relabeling functions, which will also enable us to identify equivalence classes in Lagrangian coordinates. 

\begin{definition}[Relabeling functions]
We denote by $\mathcal{G}$ the subgroup of the group of homeomorphisms $\phi$ from $\R$ to $\R$ such that 
\begin{subequations}
\begin{align}
\phi-\id \text{ and } \phi^{-1}-\id & \text{ both belong to } W^{1,\infty}(\R),\\
\text{and }\phi_\xi-1 & \text{ belongs to } L^2(\R).
\end{align}
\end{subequations}
\end{definition} 

\begin{definition}[Lagrangian coordinates]
The set $\mathcal{F}$ is composed of all quadruples of real-valued functions $(y, U, h, r)$ such that 
\begin{subequations}
\begin{align}
& (y - \id, U, h,r, y_\xi - 1, U_\xi)\in L^\infty(\R)\times [L^2(\R)\cap L^\infty(\R)]^5,\\
& y_\xi\geq 0,~h\geq 0,~y_\xi+h>0 \text{ almost everywhere on }\R, \\
\label{eqnLagrsquare} & y_\xi h=U_\xi^2+ r^2 \text{ almost everywhere on }\R, \\
& y+H\in \mathcal{G},
\end{align}
\end{subequations}
where we introduce $H$ by setting $H(\xi) = \int_{-\infty}^\xi h(s)ds$.
\end{definition}

With these definitions, we are now able to describe the transformation between the sets of Eulerian and Lagrangian coordinates. 

\begin{definition}\label{EulertoLagran}
For any $(u,\rho,\mu)$ in $\mathcal{D}$ we define $(y,U,h,r)$ by  
\begin{subequations}
\begin{align}\label{ythefirst}
 y(\xi)&=\sup\{x\in\R \mid x+\mu((-\infty,x))<\xi\},\\ \label{hthefirst}
U(\xi)& = u\circ y(\xi),\\ \label{rthefirst}
h(\xi)& = 1-y_\xi(\xi),\\
r(\xi)& = y_\xi(\xi)\,\rho\circ y(\xi).
\end{align}
\end{subequations}
Then $(y,U,h,r)$ belongs to $\F$ and we denote by $L:\mathcal{D}\mapsto\mathcal{F}$ the mapping which to any $(u,\rho,\mu)\in\mathcal{D}$ associates $(y,U,h,r)\in\mathcal{F}$ as given by \eqref{EulertoLagran}.
\end{definition}

In order to get back from Lagrangian to Eulerian coordinates, we also introduce the following mapping, where the quantity $y_\#(\nu)$ denotes the push-forward by the function $y$ of a Radon measure $\nu$ on $\R$. 

\begin{definition}
For any $ (y,U,h,r)$ in $\mathcal{F}$ we define $(u, \rho, \mu)$  by 
\begin{subequations}\label{LagEul}
\begin{align}
u(x)& =U(\xi) \text{ for any } \xi \text{ such that } x=y(\xi),\\
\mu& = y_\#(h(\xi)d\xi),\\
\rho(x)dx& = y_\#(r(\xi)d\xi).
\end{align}
\end{subequations}
Then $(u, \rho, \mu)$ belongs to $\mathcal{D}$ and we denote by $M:\mathcal{F}\mapsto \mathcal{D}$ the mapping which to any $(y,U,h,r)\in \mathcal{F}$ associates $(u, \rho, \mu)\in \mathcal{D}$ as given by \eqref{LagEul}.
\end{definition}

 We say that $X$ and $\hat{X}\in\mathcal{F}$ are equivalent, if there exists a relabeling function $\phi\in \mathcal{G}$ such that $\hat{X}=X\circ \phi$, where $X\circ \phi$ denotes $(y\circ \phi, U\circ \phi,  \phi_\xi \cdot h\circ \phi,  \phi_\xi \cdot r\circ \phi)$. 
 Upon taking equivalence classes in $\mathcal{F}$, it turns out that the mappings $L$ and $M$ are inverse to each other. 
 In particular, if we introduce the class
\begin{equation}
\mathcal{F}_0=\{X\in\mathcal{F}\mid y+H=\id \}, 
\end{equation}
then $\mathcal{F}_0$ contains exactly one representative of each equivalence class in $\mathcal{F}$.
Moreover, one readily sees that the range of the mapping $L$ is precisely the set $\mathcal{F}_0$.

The reformulation of the two-component Camassa--Holm system in Lagrangian coordinates (for conservative solutions) is given by~\eqref{eqn2CHLag} and admits a continuous semigroup of solutions. 
Denoting by $S_t(X_0)$ the solution at time $t$ with initial data $X(0)=X_0\in\mathcal{F}$, one has 
\begin{equation}
S_t(X_0\circ \phi)=S_t(X_0)\circ \phi
\end{equation}
for all $\phi\in \mathcal{G}$ (that is, the time evolution respects equivalence classes in $\mathcal{F}$).
Upon going back to Eulerian coordinates, we obtain a continuous semigroup  $M\circ S_t\circ L$ of solutions in $\mathcal{D}$ that gives rise to global conservative weak solutions of the two-component Camassa--Holm system~\eqref{eqn2CHwEul}.

% % % % % % % % % % % % % % % % % % % % % % % 
\section{Transformations of the isospectral problem}
% % % % % % % % % % % % % % % % % % % % % % % 
 
  Throughout this section, we fix some $(u,\rho,\mu)\in\mathcal{D}$ and define $\omega$ in $H^{-1}(\R)$ by 
 \begin{align}
  \omega(h) = \int_\R u(x)h(x)dx + \int_\R u_x(x)h_x(x)dx, \quad h\in H^1(\R), 
 \end{align}
 so that $\omega=u-u_{xx}$ in a distributional sense,  as well as a non-negative and finite Radon measure $\dip$ on $\R$ such that 
 \begin{align}
   \mu(B) = \int_B u_x(x)^2 dx + \dip(B) 
 \end{align}
 for every Borel set $B\subseteq\R$. 
 Let us point out that it is always possible to recover the triple $(u,\rho,\mu)$ from the distribution $\omega$ and the measure $\dip$. 
 The isospectral problem for smooth solutions of the two-component Camassa--Holm system has the form 
 \begin{align}\label{eqnDEho}
  -f_{xx} + \frac{1}{4} f = z\, \omega f + z^2 \dip f, 
 \end{align}
 where $z$ is a complex spectral parameter. 
 Moreover, there are good reasons (see \cite{ConservCH, ConservMP} as well as Section~\ref{secCI}) to expect that it also serves as an isospectral problem for global conservative solutions of the two-component Camassa--Holm system~\cite{grhora12}. 
 
 Of course, due to the low regularity of the coefficients, the differential equation~\eqref{eqnDEho} has to be understood in a distributional sense; cf.\ \cite{ConservCH, IndefiniteString, gewe14, ss03}.   
  
  \begin{definition}\label{defSolution}
  A solution of~\eqref{eqnDEho} is a function $f\in H^1_{\loc}(\R)$ such that 
 \begin{align}\label{eqnDEweakform}
   \int_{\R} f_x(x) h_x(x) dx + \frac{1}{4} \int_\R f(x)h(x)dx = z\, \omega(fh) + z^2 \int_\R f h \,d\dip 
 \end{align} 
 for every function $h\in H^1_\cc(\R)$.
 \end{definition}

 We will first show that, as long as $z$ is non-zero, the differential equation~\eqref{eqnDEho} can be transformed into an equivalent first order system of the form 
 \begin{align}\label{eqnFOSeuler}
    \begin{pmatrix} 0 & -1 \\ 1 & 0 \end{pmatrix} F_x = \begin{pmatrix} u - \frac{1}{4z} & 0 \\ 0 & 0 \end{pmatrix} F + z  \begin{pmatrix} u_x^2 & u_x \\ u_x & 1 \end{pmatrix} F + z \begin{pmatrix} \dip & 0 \\ 0 & 0 \end{pmatrix} F.
 \end{align}
 Since $\dip$ may be a genuine measure, this system has to be understood as a measure differential equation \cite{at64, be89, MeasureSL, pe88} in general: 
 A solution of the system~\eqref{eqnFOSeuler} is a function $F:\R\rightarrow\C^2$ which is locally of bounded variation with   
 \begin{align}\begin{split}\label{eqnFOSeulerInt}
   -F\big|_{x_1}^{x_2} =  \int_{x_1}^{x_2} \begin{pmatrix} 0 & 0 \\  u(s) - \frac{1}{4z} & 0 \end{pmatrix} F(s)ds & + z \int_{x_1}^{x_2} \begin{pmatrix} -u_x(s) & -1 \\ u_x(s)^2 & u_x(s) \end{pmatrix} F(s) ds \\ & + z \int_{x_1}^{x_2} \begin{pmatrix} 0 & 0 \\ 1 & 0 \end{pmatrix} F d\dip
 \end{split}\end{align}
 for all $x_1$, $x_2\in\R$. 
 In this case, the first component of $F$ is clearly locally absolutely continuous, whereas the second component is only left-continuous; cf.\ \eqref{eqnDefintmu}. 
 
 \begin{lemma}\label{lemEquCanSys}
  If the function $f$ is a solution of the differential equation~\eqref{eqnDEho}, then there is a unique left-continuous function $f^\qd$ such that 
  \begin{align}\label{eqnfqd}
   f^\qd(x) = f_x(x) - z u_x(x) f(x)
  \end{align}   
  for almost all $x\in\R$ and the function 
 \begin{align}\label{eqnFOSeulerVec}
   \begin{pmatrix} z f \\ f^\qd \end{pmatrix}   
 \end{align}
 is a solution of the system~\eqref{eqnFOSeuler}. 
 Conversely, if the function $F$ is a solution of the system~\eqref{eqnFOSeuler}, then its first component is a solution of the differential equation~\eqref{eqnDEho}. 
\end{lemma}

\begin{proof}
   Upon integrating by parts in~\eqref{eqnDEweakform}, we first note that a function $f\in H^1_\loc(\R)$ is a solution of~\eqref{eqnDEho} if and only if there is a $c\in\R$ and a constant $d\in\C$ such that 
 \begin{align}\begin{split}\label{eqnDEweakderiv}
  f_x(x)  = d +  \frac{1}{4} \int_c^x f(s)ds & -z\int_c^x u(s)f(s)+u_x(s)f_x(s)\,ds + z u_x(x)f(x) \\ &  - z^2 \int_c^x f \,d\dip
 \end{split}\end{align}
 for almost all $x\in\R$.
 So if $f$ is a solution of~\eqref{eqnDEho}, then this guarantees that there is a unique left-continuous function $f^\qd$ such that~\eqref{eqnfqd} holds for almost all $x\in\R$. 
 It is straightforward to show that the function in~\eqref{eqnFOSeulerVec} is a solution of the system~\eqref{eqnFOSeuler}. 
 
 Now suppose that $F$ is a solution of the system~\eqref{eqnFOSeuler} and denote the respective components with subscripts.  
 The first component of the integral equation~\eqref{eqnFOSeulerInt} shows that $F_{1}$ belongs to $H^1_\loc(\R)$ with 
 \begin{align*}
  z F_{2}(x) = F_{1,x}(x)  - z u_x(x) F_{1}(x) 
 \end{align*}
 for almost all $x\in\R$. In combination with the second component of~\eqref{eqnFOSeulerInt} this shows that~\eqref{eqnDEweakderiv} holds with $f$ replaced by $F_{1}$ for some $c\in\R$, $d\in\C$ and almost every $x\in\R$, which shows that $F_{1}$ is a solution of the differential equation~\eqref{eqnDEho}.  
 \end{proof}
 
 Except for the potential term (that is, the first term on the right-hand side), the equivalent first order system~\eqref{eqnFOSeuler} has the form of a canonical system; we only mention a small selection of references \cite{ardy97, dB68, hadSwi00, ka83, lema03, ro14, wi14, wiwo12}. 
 If the measure $\dip$ is absolutely continuous, then it is well known (see, for example, \cite[Section~4]{wiwo12}) that the system~\eqref{eqnFOSeuler} can be transformed (by a reparametrization) into an equivalent system with a trace normed weight matrix (that is, the matrix multiplying the spectral parameter on the right-hand side). 
 This is furthermore true in the general case upon slightly modifying the transformation; see \cite[Proof of Theorem~6.1]{IndefiniteString}. 
  In fact, upon denoting with $X=(y,U,h,r)\in\mathcal{F}_0$ the Lagrangian quantities corresponding to $(u,\rho,\mu)$ as in Definition~\ref{EulertoLagran}, it turns out that this transformation takes the system~\eqref{eqnFOSeuler} into an equivalent system of the form 
 \begin{align}\label{eqnFOSlagrange}
  \begin{pmatrix} 0 & -1 \\ 1 & 0 \end{pmatrix} G_\xi = y_\xi \begin{pmatrix} U - \frac{1}{4z} & 0 \\ 0 & 0 \end{pmatrix} G + z \begin{pmatrix} H_\xi & U_\xi \\ U_\xi & y_\xi \end{pmatrix} G.
 \end{align}
 One notes that the weight matrix is now a trace normed (that is, with trace equal to one almost everywhere) locally integrable function with determinant
 \begin{align}
  H_\xi y_\xi - U_\xi^2 = y_\xi^2\, \rho^2\circ y = r^2
 \end{align}
 in view of~\eqref{eqnLagrsquare}. 
 Thus, the system~\eqref{eqnFOSlagrange} can be understood in a standard sense.   

 \begin{lemma}\label{lemISOLag}
  Let $\sigma$ be the distribution function on $\R$ defined by  
   \begin{align}
  \sigma(x) = x + \mu((-\infty,x)), \quad x\in\R, 
%    \sigma(x) = x + \int_{-\infty}^x u_x(s)^2 ds + \int_{-\infty}^x d\dip, \quad x\in\R.  
 \end{align}
 so that $y$ is a generalized inverse of $\sigma$. 
  If the function $F$ is a solution of the system~\eqref{eqnFOSeuler}, then the function $G$ defined by 
 \begin{align}\label{eqnTransFG}
  G(\xi) = \begin{pmatrix} 1 & 0 \\ z(\sigma\circ y (\xi) - \xi) & 1 \end{pmatrix} F\circ y(\xi), \quad \xi\in\R,  
 \end{align}
 is a solution of the system~\eqref{eqnFOSlagrange}. 
 Conversely, if the function $G$ is a solution of the system~\eqref{eqnFOSlagrange}, then~\eqref{eqnTransFG} defines a function $F$ that is a solution of the system~\eqref{eqnFOSeuler}. 
 \end{lemma}
 
 \begin{proof}
 To begin with,  let us note the simple identities 
 \begin{align*}
  y \circ \sigma(x) & = x, \quad x\in\R; & \sigma \circ y(\xi) & =  \xi, \quad \xi\in\ran{\sigma}. 
 \end{align*}
 In particular,  since $y$ is locally constant on $\R\backslash\ran{\sigma}$, this gives the equality 
 \begin{align}\label{eqnLTxi}
  y_\xi(\xi) G(\xi) = y_\xi(\xi) F\circ y(\xi)
 \end{align}
 for almost all $\xi\in\R$, if $F$ and $G$ are related by~\eqref{eqnTransFG}.
% More precisely, we clearly have $G(\xi) = F\circ y(\xi)$ when $\xi$ lies in $\ran{\sigma}$.  
% Now note that $\R\backslash\ran{\sigma}$ coincides with the union $\cup_{x\in I} (\sigma(x),\sigma(x+)]$ where $I\subseteq\R$ is the (countable) set of points of discontinuity of $\sigma$.
% We may ignore the right endpoints since they form a null set and it remains to note that $y$ is constant on each interval $(\sigma(x),\sigma(x+))$.  
 The remaining ingredients are two substitution formulas:
 Firstly, for every function $h\in L^1_{\loc}(\R)$, we have 
 \begin{align}\label{eqnIntpartI}
  \int_{y(\xi_1)}^{y(\xi_2)} h(s) ds = \int_{\xi_1}^{\xi_2} y_\xi(s) h\circ y(s) ds, \quad \xi_1,\, \xi_2\in\R,
 \end{align}
 according to, for example, \cite[Corollary~5.4.4]{bo07}. 
 Secondly, we will also use the identity   
 \begin{align}\label{eqnIntpartII}
  \int_{\sigma(x_1)}^{\sigma(x_2)} h\circ y(s)ds = \int_{x_1}^{x_2} h(s) ds + \int_{x_1}^{x_2} h(s)d\mu(s), \quad x_1,\, x_2\in\R, 
 \end{align}
 which holds for all continuous functions $h$ on $\R$ (see, for example \cite[Theorem~3.6.1]{bo07})
% Use $X=Y=\R$ with the Lebesgue measure on $X$. Then its image under $y$ is the measure $\sigma_x$ since $y^{-1}[x_1,x_2) = [\sigma(x_1),\sigma(x_2))$. It remains to apply \cite[Theorem~3.6.1]{bo07}. 

 Now suppose that $F$ is a solution of the system~\eqref{eqnFOSeuler} and let $\xi_1$, $\xi_2\in\R$.
 Then the integral equation~\eqref{eqnFOSeulerInt}, identity~\eqref{eqnLTxi} as well as~\eqref{eqnIntpartI} and~\eqref{eqnIntpartII} give 
 \begin{align*}\begin{split}
   -F\big|_{y(\xi_1)}^{y(\xi_2)} =  \int_{\xi_1}^{\xi_2} y_\xi(s) \begin{pmatrix} 0 & 0 \\  U(s) - \frac{1}{4z} & 0 \end{pmatrix} G(s)ds & + z \int_{\xi_1}^{\xi_2} \begin{pmatrix} -U_\xi(s) & -y_\xi(s) \\ -y_\xi(s) & U_\xi(s) \end{pmatrix} G(s) ds \\ & + z \int_{\sigma\circ y(\xi_1)}^{\sigma\circ y(\xi_2)} \begin{pmatrix} 0 & 0 \\ 1 & 0 \end{pmatrix} G(s) ds.
 \end{split}\end{align*} 
  Moreover, since $y$ is constant on $[\sigma\circ y(\xi_i),\xi_i]$ when $\xi_i\not\in\ran{\sigma}$ we have 
  \begin{align*}
    z (\xi_i - \sigma\circ y(\xi_i)) F_1\circ y(\xi_i) = z \int_{\sigma\circ y(\xi_i)}^{\xi_i} F_1\circ y(s)ds = z \int_{\sigma\circ y(\xi_i)}^{\xi_i} G_1(s)ds
  \end{align*}
  for $i=1$, $2$.   
  After a straightforward calculation, all this finally gives 
 \begin{align}\begin{split}\label{eqnFOSlagrangeInt}
   -G\big|_{\xi_1}^{\xi_2} =  \int_{\xi_1}^{\xi_2} y_\xi(s) \begin{pmatrix} 0 & 0 \\  U(s) - \frac{1}{4z} & 0 \end{pmatrix} G(s)ds & + z \int_{\xi_1}^{\xi_2} \begin{pmatrix} -U_\xi(s) & -y_\xi(s) \\ H_\xi(s) & U_\xi(s) \end{pmatrix} G(s) ds,
 \end{split}\end{align}
  which shows that $G$ is a solution of the system~\eqref{eqnFOSlagrange}. 
  
  For the converse, suppose that $G$ is a solution of the system~\eqref{eqnFOSlagrange}. 
  In order to show that a function $F$ is well-defined by~\eqref{eqnTransFG}, let $\xi_1$, $\xi_2\in\R$ such that $y(\xi_1) = y(\xi_2)$ and assume that $\xi_1\leq\xi_2$ without loss of generality. 
  Then the functions $y$ and $U$ are constant on the interval $[\xi_1,\xi_2]$ and we readily infer that 
  \begin{align*}
   G_1(\xi_1) & = G_1(\xi_2),  & -G_2\big|_{\xi_1}^{\xi_2} & =  z (\xi_2-\xi_1) G_1(\xi_1),
  \end{align*}
 which shows that $F$ is well-defined (also note that the range of $y$ is all of $\R$). 
 Now for any given $x_1$, $x_2\in\R$, we may evaluate 
 \begin{align*}
  -F\big|_{x_1}^{x_2} = -G\big|_{\sigma(x_1)}^{\sigma(x_2)}
 \end{align*}
  using the integral equation~\eqref{eqnFOSlagrangeInt}, identity~\eqref{eqnLTxi} as well as~\eqref{eqnIntpartI} and~\eqref{eqnIntpartII} to see that $F$ is a solution of the system~\eqref{eqnFOSeuler}. 
 \end{proof}
 
 Note that the matrix in~\eqref{eqnTransFG} disappears if the measure $\dip$ is absolutely continuous and the transformation in Lemma~\ref{lemISOLag} becomes a simple reparametrization. 
 
 In order to show what relabeling means in terms of the isospectral problem, let $\phi\in\mathcal{G}$ be a relabeling function and consider the relabeled Lagrangian variables $\hat{X}=X\circ\phi$ as in Section~\ref{secEtoL}. 
 Then it is not hard to see that the original system~\eqref{eqnFOSlagrange} is indeed equivalent to the corresponding system with relabeled variables 
  \begin{align}\label{eqnFOSlagrangeRL}
  \begin{pmatrix} 0 & -1 \\ 1 & 0 \end{pmatrix} \hat{G}_\xi = \hat{y}_\xi \begin{pmatrix} \hat{U} - \frac{1}{4z} & 0 \\ 0 & 0 \end{pmatrix} \hat{G} + z  \begin{pmatrix} \hat{H}_\xi & \hat{U}_\xi \\ \hat{U}_\xi & \hat{y}_\xi \end{pmatrix} \hat{G},
 \end{align}
 by means of the following simple transformation. 
 
 \begin{lemma}\label{lemISOrl}
  A function $G$ is a solution of the system~\eqref{eqnFOSlagrange} if and only if the function $\hat{G}=G\circ\phi$  is a solution of the system~\eqref{eqnFOSlagrangeRL}. 
 \end{lemma}
 
 \begin{proof}
  The claim follows immediately upon applying the substitution rule \cite[Corollary~5.4.4]{bo07} to the integral equation~\eqref{eqnFOSlagrangeInt}. 
 \end{proof}
 
Concluding, we have seen that~\eqref{eqnFOSeuler} and~\eqref{eqnFOSlagrange} represent the same (that is, equivalent up to reparametrizations) first order system with different kinds of normalizations. 
The Lagrangian coordinates in $\mathcal{F}_0$ correspond to trace normalization of the weight matrix, whereas the Eulerian coordinates in $\mathcal{D}$ correspond to normalization of the bottom-right entry in the weight matrix.

 % % % % % % % % % % % % % % % % % % 
\section{Complete integrability}\label{secCI}
 % % % % % % % % % % % % % % % % % % 

 In this final section, we will show that the Lagrangian version~\eqref{eqn2CHLag} of the two-component Camassa--Holm system is completely integrable in the sense that it can be formulated as the condition of compatibility for an overdetermined linear system. 
  To this end, let us first write down the corresponding reformulation of the two-component Camassa--Holm system~\eqref{eqn2CHwEul} as a condition of compatibility: 
 For sufficiently smooth functions $u$, $\rho$ and $p$, we set  
  \begin{align}
  V & = \begin{pmatrix} 0 & 0 \\  \frac{1}{4z}-u & 0 \end{pmatrix} -z \begin{pmatrix} -u_x & -1 \\ u_x^2 + \rho^2 &u_x \end{pmatrix}, \\
  W & = \frac{1}{4z} \begin{pmatrix} 0 & 0 \\ u -\frac{1}{2z} & 0 \end{pmatrix} + \begin{pmatrix} 0 &- \frac{1}{2} \\ p  & 0 \end{pmatrix} +z u \begin{pmatrix} -u_x & -1 \\ u_x^2+ \rho^2 & u_x \end{pmatrix}.
 \end{align}
 Then a straightforward calculation shows that 
 \begin{align}
     V_t & = - \begin{pmatrix} 0 & 0 \\ u_t & 0 \end{pmatrix} - z  \begin{pmatrix} -u_{xt} & 0 \\ 2u_x u_{xt} + 2\rho\rho_t  & u_{xt} \end{pmatrix},  \\
 \begin{split} W_x & = \frac{1}{4z} \begin{pmatrix} 0 & 0 \\  u_x  & 0 \end{pmatrix} + \begin{pmatrix} 0 & 0 \\ p_x & 0 \end{pmatrix} \\
     &\qquad + z u_x \begin{pmatrix} -u_x & -1 \\ u_x^2 + \rho^2 & u_x \end{pmatrix} + z u \begin{pmatrix} -u_{xx} & 0 \\ 2u_x u_{xx} + 2\rho\rho_x & u_{xx} \end{pmatrix}, \end{split} \\
 \begin{split}  [V,W] & = \frac{1}{4z} \begin{pmatrix} 0 & 0 \\ u_x & 0 \end{pmatrix}  - \begin{pmatrix} 0 & 0 \\ uu_x & 0 \end{pmatrix} \\
     &\qquad + z \left(u^2-p\right) \begin{pmatrix} -1  & 0 \\ 2u_x & 1 \end{pmatrix} - \frac{z}{2} \begin{pmatrix} u_x^2+\rho^2   & 2u_x \\ 0 & -\left(u_x^2 + \rho^2\right) \end{pmatrix}. \end{split}
 \end{align}
 Upon collecting equal powers of $z$, we see that the condition of compatibility 
 \begin{align}
   V_t-W_x + [V,W] = 0
 \end{align}
  for the overdetermined system
   \begin{align}\label{eqnCCeuler}
  \Psi_x & = V \Psi, & \Psi_t & = W \Psi, 
 \end{align} 
% becomes 
% \begin{align}
%  V_t-W_x + [V,W] & = - \begin{pmatrix}  0 & 0\\ u_t+uu_x+P_x & 0 \end{pmatrix}\\
%  & \quad - z \begin{pmatrix} - u_{xt}& 0\\Ê2u_xu_{xt}+ 2\rho\rho_t & u_{xt}\end{pmatrix}\\
%  & \quad +z \begin{pmatrix} u_x^2+uu_{xx}+p-u^2-\frac{1}{2} u_x^2-\frac12\rho^2 & 0\\
%   -u_x^3 - u_x\rho^2 - 2uu_xu_{xx} - 2u\rho\rho_x + 2u^2u_x - 2u_x p  & -u_x^2-uu_{xx}-p+u^2+\frac12 u_x^2+\frac12\rho^2 \end{pmatrix} = 0,
% \end{align}
% which 
is indeed equivalent to the two-component Camassa--Holm system~\eqref{eqn2CHwEul}. 
%\begin{align}\begin{split}
% u_t+uu_x+p_x & =0,\\ 
%  \rho_t + (u\rho)_x & = 0, 
%\end{split}\end{align}
%where $P$ is related to $u$ and $\rho$ via 
%\begin{align}
%p-p_{xx} & =u^2+\frac12 u_x^2+\frac12\rho^2.
%\end{align}

Now suppose that the functions $y$, $U$, $H$ and $P$ are sufficiently smooth and define
\begin{align}
N & = y_\xi \begin{pmatrix} 0 & 0 \\ \frac{1}{4z} -U & 0\end{pmatrix}-z\begin{pmatrix} -U_\xi & -y_\xi \\ H_\xi & U_\xi \end{pmatrix}, \\
M & = \frac{1}{2z} \begin{pmatrix} 0 & 0 \\ U - \frac{1}{4z} & 0 \end{pmatrix} - \begin{pmatrix} 0 & \frac12 \\  U^2 - P  & 0\end{pmatrix}.
\end{align}
Then one readily computes that 
\begin{align}
 N_t& =\frac{1}{4z}\begin{pmatrix} 0 & 0\\ y_{\xi t}& 0\end{pmatrix}   - \begin{pmatrix}0 & 0\\ U_t y_\xi + U y_{\xi t} & 0\end{pmatrix}   - z\begin{pmatrix}Ê- U_{\xi t} & - y_{\xi t}\\ H_{\xi t} &  U_{\xi t}\end{pmatrix}, \\
 M_\xi & = \frac{1}{2z} \begin{pmatrix} 0 & 0\\ U_\xi & 0\end{pmatrix}  - \begin{pmatrix} 0 & 0\\ 2UU_\xi - P_\xi & 0\end{pmatrix}, \\ 
 \begin{split}
[N, M] & = \frac{1}{4z} \begin{pmatrix} 0 & 0\\ U_\xi & 0\end{pmatrix} - \begin{pmatrix} 0 & 0 \\ UU_\xi & 0\end{pmatrix}Ê\\ 
& \qquad +z \left(U^2-P\right) \begin{pmatrix} -y_\xi & 0 \\Ê2U_\xi & y_\xi \end{pmatrix} - \frac{z}{2} \begin{pmatrix}  H_\xi & 2U_\xi \\Ê0 &  -H_\xi \end{pmatrix}.
 \end{split}
\end{align} 
Again, upon collecting equal powers of $z$, the condition of compatibility 
\begin{align} 
   N_t - M_\xi +[N, M]=0
\end{align}
  for the overdetermined system
   \begin{align}
  \Psi_\xi & = N \Psi, & \Psi_t & = M \Psi, 
 \end{align} 
% becomes 
% \begin{align}
%  N_t- M_\xi +[N, M]=& \frac{1}{4z} \begin{pmatrix} 0 & 0 \\Êy_{\xi t}-U_\xi & 0\end{pmatrix} \\ \nonumber
%& \quad + \begin{pmatrix} 0 & 0 \\ -U_ty_\xi-Uy_{\xi t}+UU_\xi-P_\xi & 0 \end{pmatrix} \\ \nonumber
%& \quad + z \begin{pmatrix} U_{\xi t}-\frac12 H_\xi -(U^2-P)y_\xi & y_{\xi t}-U_\xi\\ -H_{\xi t}+2(U^2-P)U_\xi & -U_{\xi t}+\frac12 H_\xi +(U^2-P)y_\xi\end{pmatrix} \\  \nonumber
%& \quad = \begin{pmatrix} 0 & 0\\ 0 & 0\end{pmatrix},
%\end{align}
% which 
 is seen to be equivalent to the system 
\begin{subequations}\label{eq}
\begin{align}
 \label{eqnZC1} y_{\xi t}& =U_\xi, \\
\label{eqnZC3} U_{\xi t}& = \frac12 H_\xi +(U^2-P)y_\xi,\\
\label{eqnZC4} H_{\xi t} & = 2(U^2-P)U_\xi, \\
\label{eqnZC2} U_ty_\xi & = -P_\xi.
\end{align}
\end{subequations}
Note that in this case, the function $P$ is determined by $y$, $U$, $H$ to the extent that 
\begin{align}\label{eqnDEP}
 y_\xi P_{\xi\xi} - y_{\xi\xi} P_\xi - y_\xi^3 P = -\frac{1}{2}y_\xi^2 (2U^2y_\xi + H_\xi ),
\end{align}
which follows upon combining~\eqref{eqnZC2} and~\eqref{eqnZC3}. 
We finally want to show that this system indeed reduces to the system~\eqref{eqn2CHLag} under the additional assumption that the functions $y-\id$, $U$, $H$ and $P$ (as well as their derivatives) decay spatially. 
In fact, it is immediate that the system~\eqref{eqn2CHLag} implies~\eqref{eq}. 
Conversely, given a solution of the system~\eqref{eq}, we first obtain~\eqref{eqn2CHLag1} from~\eqref{eqnZC1} upon exploiting the decay assumption. 
Furthermore, one sees that for every $t\in\R$ we have 
\begin{equation}
 P(\xi,t)=\frac14 \int_\R e^{-\vert y(\xi,t)-y(s,t)\vert}(2U^2y_\xi+h)(s,t)ds, \quad \xi\in\R,
\end{equation}
since the function on the right-hand side in this equation is a solution of~\eqref{eqnDEP} as well (also take into account that both functions in this equation are constant whenever $y$ is constant). 
In view of~\eqref{eqnZC4}, this shows that~\eqref{eqn2CHLag3} holds with $P$ given by~\eqref{intrepP3}. 
Because we clearly have 
\begin{align}
 P_\xi(\xi,t) = y_\xi(\xi,t) Q(\xi,t), \quad \xi\in\R,
\end{align}
with $Q$ defined as in~\eqref{eqnDefQ}, we see from~\eqref{eqnZC2} that~\eqref{eqn2CHLag2} holds whenever $y_\xi$ is non-zero.  
If $y$ is constant on some interval, then~\eqref{eqnZC3} and~\eqref{eqnDefQ} show that 
\begin{align}
 U_{t\xi} = \frac12 H_\xi = -Q_\xi,
\end{align}
which implies that~\eqref{eqn2CHLag2} holds everywhere indeed. 
It remains to note that the time evolution of $r$ can be derived by using $r^2 = y_\xi h - U_\xi^2$, which yields $2rr_t=0$.

%\bigskip
%\noindent
%{\bf Acknowledgments.}

\end{document}